\documentclass[epj,final]{svjour}
\usepackage{epsfig}
\usepackage{amsmath,amssymb}
\begin{document}
\def\vm{v_{max}}
\def\q{\bar{p}}
\newcommand{\btau}{\boldsymbol{\tau}}
\newcommand{\bttau}{\boldsymbol{\tilde{\tau}}}
\newcommand{\btaujn}{\boldsymbol{\tau}_j^{(n)}}
\newcommand{\balpha}{\boldsymbol{\alpha}}
\newcommand{\bbeta}{\boldsymbol{\beta}}
\newcommand{\nn}{{\cal N}}
\newcommand{\Pt}{\tilde{\mathbf P}}
\newcommand{\cP}{{\cal P}}
\newcommand{\be}{\begin{equation}}
\newcommand{\ee}{\end{equation}}
\newcommand{\bea}{\begin{eqnarray}}
\newcommand{\eea}{\end{eqnarray}}
\newcommand{\nonu}{\nonumber\\}


\title{\bf The Nagel-Schreckenberg model revisited}

\author{Andreas Schadschneider}

\institute{Institut f\"ur Theoretische Physik,
  Universit\"at zu K\"oln,
  D--50937 K\"oln, Germany,
  email: {\tt as@thp.uni-koeln.de}}

\date{\today}


\abstract{
The Nagel-Schreckenberg model is a simple cellular automaton for a
realistic description of single-lane traffic on highways. 
For the case $\vm=1$ the
properties of the stationary state can be obtained exactly. For the more
relevant case $\vm>1$, however, one has to rely on Monte Carlo simulations
or approximative methods. Here we study several analytical approximations and
compare with the results of computer simulations. The role of the braking
parameter $p$ is emphasized. 
It is shown how the local structure of the stationary state depends on
the value of $p$. This is done by combining the results of computer
simulations with those of the approximative methods.}


\maketitle

\section{Introduction}

Cellular automata (CA) do not only serve
as simple model systems for the investigation of problems in statistical
mechanics, but they also have numerous applications to 'real' problems
\cite{Wolfram}. Therefore it is not surprising that in recent years CA have 
become quite popular for the simulation of traffic flow (see e.g.\ 
\cite{juelich,tgf97}).

CA are -- by design -- ideal for large-scale computer simulations.
On the other hand, analytical approaches for the description of
CA are notoriously difficult. This is mainly due to the discreteness
and the use of a parallel updating scheme (which introduces 'non-locality'
into the dynamics). In addition, these models are defined through
dynamical rules (e.g.\ transition probabilities) and usually one does
not have a 'Hamiltonian' description. Therefore standard methods
are not applicable. Furthermore, one has to deal with systems
which do not satisfy the detailed balance condition.

However, there is a need for exact solutions or, at least, for good
approximations. These results as well as other exact statements may
help to greatly reduce the need for computer resources.  The
interpretation of simulation data is often difficult because of the
'numerical noise' and finite-size effects. Even in the cases where an
exact solution is not possible, a combination of analytical and
numerical methods might provide better insights. This is especially
true for non-equilibrium systems where only a few exact or general
results exist which could serve as a guideline.

In recent years, several analytical approximation methods for the
Nagel-Schreckenberg model \cite{NagelS} for single-lane highway traffic 
have been proposed \cite{ss93,SSNI,COMF,eden}. All of
these approximations yield the exact result for the stationary state
in the case $\vm=1$. However, these investigations focused on the 
so-called fundamental diagram, i.e.\ the flow-density relation. Here we 
will reinvestigate these approximations and calculate further quantities
of interest in order to determine the accuracy. We will also discuss
several limits and emphasize the effect of the braking probability $p$.
We focus on the effect of $p$ on the microscopic structure of the
stationary state. Such microscopic characterizations have recently been
used successfully for the asymmetric exclusion process \cite{ernst,kolo}.
Apart from $\vm=1$ here only the case of $\vm=2$ is investigated
since one does not expect a qualitively different behaviour for $\vm>2$
\cite{SSNI}.


The paper is organized as follows: First we will briefly recall the 
definition of the Nagel-Schreckenberg model in Sec.\ \ref{sec_nasch} and
the different analytical approximations in Sec.\ \ref{sec_approx}.
In Sec.\ \ref{sec_phys} several physical quantities, e.g.\ the
fundamental diagram, headway and jam-size distributions and correlation
lengths, are calculated using the analytical results.
In Sec.\ \ref{sec_struct} the predictions of the approximations
are compared with each other and with results
from computer simulations. 
The local structure of the stationary state is investigated as a 
function of the randomization $p$.
In the final Sec.\ \ref{sec_summary} a summary of the results together
with our conclusion are given.


\section{The Nagel-Schreckenberg model}
\label{sec_nasch}

The Nagel-Schreckenberg (NaSch) model \cite{NagelS} is a probabilistic
cellular automaton. Space and time are discrete and hence also
the velocities. The road is divided into cells.
The length of a cell is determined by the front-bumper to front-bumper
distance of cars in the densest jam and is usually taken to be 7.5~m.
Each cell can either be empty or occupied by just one car. The state
of car $j$ ($j=1,\ldots,N$) is characterised by an internal 
parameter $v_j$ ($v_j=0,1,\ldots, \vm$), the instantaneous velocity of the 
vehicle. In order to obtain the
state of the system at time $t+1$ from the state at time $t$, one has
to apply the following four rules to all cars at the same time
(parallel dynamics) \cite{NagelS}:
\begin{description}
\item[{\bf R1}] Acceleration:\ \ \  $v_j(t+1/3)=\min(v_j(t)+1,\vm)$
\item[{\bf R2}] Braking:\ \ \  $v_j(t+2/3)=\min(d_j(t),v_j(t+1/3))$ \ \ \ 
\item[{\bf R3}] Randomization:\ \ \  $v_j(t+1) {\stackrel{p}{=}}\
\max(v_j(t+2/3)-1,0)$\ \ \ 
\phantom{Randomization:\ \ \  }\ with probability $p$
\item[{\bf R4}] Driving:\ \ \  car $j$ moves $v_j(t+1)$ cells.
\end{description}
Here $d_j(t)$ denotes the number of empty cells in front of car $j$, i.e.\ 
the so-called headway. For $\vm=5$ a calibration of the model shows
that each timestep $t\to t+1$ corresponds to approximately 1 sec in 
real time \cite{NagelS}. 
\begin{figure}[ht]
\centerline{\epsfig{figure=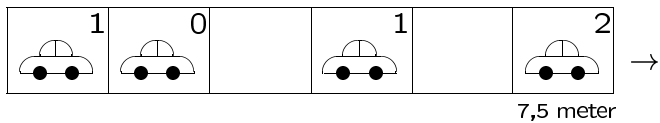,height=1.5cm}}
\caption{Configuration in the Nagel-Schreckenberg model. The number in the 
upper right corner gives the velocity of the car.}
\label{fig_conf}
\end{figure}
For simplicity we will consider only periodic boundary conditions so 
that the number of cars is conserved.
The maximum velocity $\vm$ can be interpreted as a speed limit and is
therefore taken to be identical for all cars. Fig.\ \ref{fig_conf}
shows a typical configuration. Throughout the paper we will assume that
the cars move from left to right.

The four steps have simple interpretations. Step R1 means that every driver
wants to drive as fast as possible or allowed. Step R2 avoids crashes
between the vehicles. The randomization step R3 takes into account several
effects, e.g.\ road conditions (e.g.\ slope, weather) or psychological 
effects (e.g.\ velocity fluctuations in free traffic).
An important consequence of this step is the introduction of overreactions 
at braking which are crucial for the occurance of spontaneous jam
formation. Finally, step R4 is the actual motion of the vehicles.

The NaSch model is a minimal model in the sense that all four steps R1-R4
are necessary to reproduce the basic properties of real traffic. For more 
complex situations (e.g. 2-lane traffic \cite{2lane} or city 
traffic \cite{city}) additional rules have to be formulated. 


\section{Analytical Methods}
\label{sec_approx}

In this section several analytical approaches which have been used
for the description of the NaSch model are reviewed. The simplest,
a mean-field (MF) theory, completely neglects correlations. Since MF theory
turned out to be inadequate even for the fundamental diagram, improved
methods have been developed which allow to take into account short-range
correlations exactly. All methods described here are {\em microscopic} 
theories since macroscopic theories are not able to
describe the NaSch model properly. However, they are extremely useful
and accurate for special variants of the NaSch model, e.g.\ the VDR model
\cite{robert} in the slow-to-start limit.

In the following the analytical approaches are discussed briefly for the
cases $\vm=1$ and $\vm=2$. All methods other than MF theory are
exact for $\vm=1$. For $\vm=2$, on the other hand, they are only 
approximations.

Applications of the analytical methods to the calculation of physical
quantities, e.g.\ the fundamental diagram, jam-size distributions and
correlation lengths, are given in Sec.\ \ref{sec_phys}.


\subsection{Mean-Field Theory}
\label{sec_MF}

The simplest analytical approach to the NaSch model is a (microscopic)
mean-field (MF) theory \cite{SSNI}. Here one considers the density 
$c_v(j,t)$ of cars with velocity $v$ at site $j$ and time $t$. In the MF 
approach, correlations between sites are completely neglected.

For $\vm=1$ the MF equations for the stationary state ($t\to\infty$)
read \cite{SSNI}:
\begin{eqnarray}
c_0 &=& (c+pd)c\ ,\label{mf1}\\
c_1 &=& \q cd \label{mf2}
\end{eqnarray}
with $c=c_0+c_1$, $d=1-c$ and $\q =1-p$. 
The flow is simply given by $f_{MF}(c)=c_1$.

For random-sequential dynamics\footnote{In random-sequential dynamics in 
each timestep a cell which is updated is picked at random.} the MF approach 
is known to be exact for $\vm=1$ \cite{NagelS}. For parallel dynamics, 
however, MF theory  underestimates the flow considerably (see 
Sec.\ \ref{sec_funda}).

For $\vm=2$ the rate equations for the densities are given by \cite{SSNI}
\bea
c_0 &=& (c+pd)c_0 + (1+pd)c(c_1+c_2)\ ,\label{mf2a}\\
c_1 &=& d\left[\q c_0 + (\q c+pd)(c_1+c_2)\right]\ ,\label{mf2b}\\
c_2 &=& \q d^2(c_1+c_2)\ .\label{mf2c}
\eea
The solution is given by
\bea
c_0 &=&\frac{(1+pd)c^2}{1-pd^2}\ ,\nonumber\\
c_1 &=&\frac{\q (1-\q d^2)dc}{1-pd^2}\ ,\\
c_2 &=&\frac{\q ^2d^3c}{1-pd^2}\ ,\nonumber
\eea
and the flux can be calculated using $f_{MF}(c)=c_1+2c_2$.
Again the flow is underestimated considerably (see Sec.\ \ref{sec_funda}).
This is true for arbitrary $\vm$. The general form of the MF equations
can be found in \cite{SSNI}.


\subsection{Cluster Approximation}
\label{sec_clust}

The cluster approximation \cite{ss93,SSNI} is a systematic improvement of 
MF theory which
takes into account short-ranged correlations between the cells.
In the $n$-cluster approximation a cluster of $n$ neighbouring cells is
treated exactly. The cluster is then coupled 
to the rest of the system in a self-consistent way. Related approximations
have already been used (under different names) for other models
\cite{kikuchi,guto,benA,cris}.

In order to simplify the description, we sometimes choose a slightly different
update-ordering  R2-R3-R4-R1 instead of R1-R2-R3-R4, i.e.\ we look at the 
system after  the acceleration step. Then there are no cars with $v=0$ and
effectively we have to deal with one equation less. In the following
we will use occupation variables $\tau_j$ where $\tau_j=0$, if cell $j$ is
empty, and $\tau_j=v$, if cell $j$ is occupied by a car with velocity $v$.
The change of the ordering has to be taken into account in the calculation
of observables. The flow is given by $f(c)=\q P(1,0)$.

For an $n-$cluster of $n$ consecutive cells with state variables
$\btaujn(t)=(\tau_j(t),\ldots,\tau_{j+n-1}(t))$ at time $t$ it is 
straightforward to derive an exact evolution equation for the cluster 
probability $P(\tau_j(t), \ldots,\tau_{j+n-1}(t))$. One has to take into 
account that cars can enter the cluster from one of the $\vm$ cells to the 
left of the cluster and can leave the cluster to one of the $\vm$ cells to the
right. Therefore the state $\btaujn(t+1)$ of a cluster depends not only 
on its state $\btaujn$ at time $t$, but also on the 
neighbouring cells $(\tau_{j-\vm}(t),\ldots,$ $\tau_{j-1}(t))$ and 
$(\tau_{j+n}(t),\ldots,$ $\tau_{j+n+\vm-1}(t))$.

In the stationary state the master equation for cluster
$\btau^{(n)} =(\tau_j,\ldots,\tau_{j+n-1})$ of $n$ cells 
has the following structure:
\begin{equation}
P(\btau^{(n)})=\sum_{{\btau^{(n+2\vm)}}} 
W(\btau^{(n+2\vm)} \to\btau^{(n)})P(\btau^{(n+2\vm)})
\label{master_struct}
\end{equation}
where $\btau^{(n+2\vm)}=(\tau_{j-\vm},\ldots,\tau_{j+n+\vm-1})$. 
The transition probabilities $W(\btau^{(n+2\vm)}\to\btau^{(n)})$ have to 
be determined from the rules R1--R4.
Note that the master equation for $n-$clusters involves 
$(n+2\vm)-$clusters. 
If the stationary state is translation-invariant
the probabilities $P(\tau_j \ldots,\tau_{j+n-1})$ are
independent of $j$.

\begin{figure}[ht]
\centerline{\psfig{figure=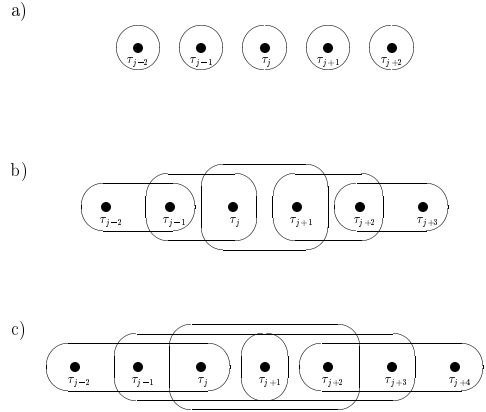,bbllx=115pt,bblly=310pt,bburx=570pt,bbury=715pt,height=6cm}}
\caption{Graphical representation of the $n-$cluster approximation for
a) $n=1$ (i.e.\ mean-field theory), b) $n=2$, and c) $n=3$. Shown is the 
central $n-$cluster starting at site $j$ and all clusters which have to
be taken into account in the master equation (\ref{master_struct}) 
for $\vm=2$.}
\label{fig_3cluster}
\end{figure}

In order to obtain a closed set of equations one has to express the
$(n+2\vm)-$clusters through the $n-$cluster probabilities. At this
point some approximation has to be made. Usual one uses a factorization
into products of $n-$clusters. We illustrate this for $\vm=2$ and
$n=3$ (see Fig.\ \ref{fig_3cluster}):
\begin{eqnarray}
P(\btau^{(7)})&=&P(\tau_{j-2}|\underline{\tau_{j-1},\tau_{j}})\cdot 
P(\tau_{j-1}|\underline{\tau_{j},\tau_{j+1}})\cdot 
\nonumber \\
&\cdot& P(\tau_{j},\tau_{j+1},\tau_{j+2})\cdot\\
&\cdot& P(\underline{\tau_{j+1},\tau_{j+2}}|
\tau_{j+3})\cdot
P(\underline{\tau_{j+2},\tau_{j+3}}|\tau_{j+4})
\nonumber
\end{eqnarray}
with the conditional probabilities
\begin{eqnarray}
P(\tau_1|\underline{\tau_2,\ldots,\tau_n})&=&
\frac{P(\tau_1,\ldots,\tau_n)}{\sum_{\tau} P(\tau,\tau_2,\ldots,\tau_n)}
\ , \label{condprob1}\\
P(\underline{\tau_1,\ldots,\tau_{n-1}}|\tau_n)&=&
\frac{P(\tau_1,\ldots,\tau_n)}{\sum_{\tau}
P(\tau_1,\ldots,\tau_{n-1},\tau)}\ .\label{condprob2}
\end{eqnarray}

If we denote the probability to find the system in a configuration
$(\tau_1,\ldots,\tau_L)$ by $P(\tau_1,\ldots,\tau_L)$ the 1-cluster 
approximation means a simple factorization 
\begin{equation}
P(\tau_1,\ldots,\tau_L)=\prod_{j=1}^L P(\tau_j)\ .
\end{equation}
This is nothing but the mean-field theory of section \ref{sec_MF}.
For the 2-cluster approximation one has a factorization of the form
\begin{eqnarray}
P(\tau_1,\ldots,\tau_L)&\propto& P(\tau_1,\tau_2)P(\tau_2,\tau_3)
\cdots P(\tau_{L-1},\tau_L)P(\tau_L,\tau_1)\ .\nonumber\\
& &\phantom{P(\tau_{L-1},\tau_L)P(\tau_L,\tau_1)}
\label{2clusterapp}
\end{eqnarray}
The 3-cluster approximation is depicted graphically in Fig.\
\ref{fig_3cluster}c. 

In general, the master equation in $n$-cluster approximation leads to 
$(\vm+1)^n$ nonlinear equations. This number can be reduced by using
the so-called Kolmogorov consistency conditions \cite{guto}
\begin{eqnarray}
\sum_{\tau=0}^{\vm} P(\tau_1,\ldots,\tau_{n-1},\tau)
&=&P(\tau_1,\ldots,\tau_{n-1})\nonumber\\
=\sum_{\tau=0}^{\vm} &P&(\tau,\tau_1,\ldots,\tau_{n-1})\ .
\end{eqnarray}
Nevertheless, a solution is only feasible
for relatively small cluster-sizes \cite{SSNI}. The quality of the
approximation improves with increasing $n$ and for $n\to\infty$ the
$n$-cluster result becomes exact. 

However, for $v_{max}=1$ already the 2-cluster approximation is exact 
\cite{ss93,SSNI}. The 2-cluster probablities for
the stationary state are given explicitly by
\begin{eqnarray}
P(0,0)&=&1-c-P(1,0)\  ,\nonumber\\
P(1,1)&=&c-P(1,0)\  ,\label{2clustsol1}\\
P(1,0)&=&P(0,1)=\frac{1}{2\q }\left[1-\sqrt{1-4\q c(1-c)}\right]\  ,
\nonumber
\end{eqnarray}
where again $\q =1-p$. 
The flow is given by $f(c)=\q P(1,0)$.

Higher order cluster approximations with $n>2$ yield the same result 
(\ref{2clustsol1}). This indicates that (\ref{2clustsol1}) is exact. Indeed
this has been proven in \cite{SSNI} by a combinatorial argument.

\begin{figure}[t]
\centerline{\psfig{figure=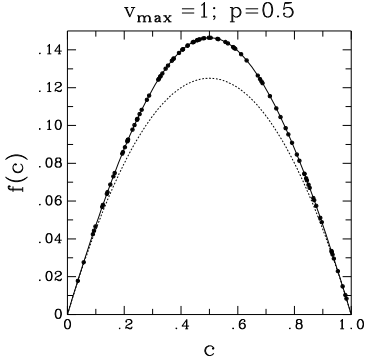,bbllx=40pt,bblly=165pt,bburx=565pt,bbury=680pt,height=5cm}}
\caption{Fundamental diagram for $\vm=1$: Comparison of computer simulations
($\bullet$) with the exact solution (full line) and the mean-field
result (broken line).}
\label{fig_vmax1}
\end{figure}

\begin{figure}[t]
\centerline{\psfig{figure=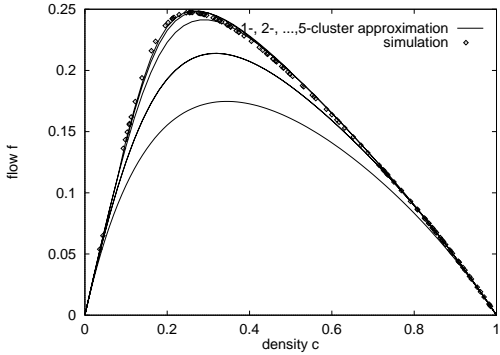,bbllx=40pt,bblly=210pt,bburx=535pt,bbury=560pt,height=5cm}}
\caption{Comparison of simulation results with the $n$-cluster approximation
($n=1,\ldots,5$ from bottom to top) for the fundamental diagram for $\vm=2$ 
and $p=1/2$.}
\label{fig_fund2}
\end{figure}

For $\vm=2$ already the 2-cluster approximation yields a nonlinear
system of equations which has to be solved numerically.
The fundamental diagrams obtained from the $n$-cluster
approximation ($n=1,\ldots,5$) are compared in 
Fig.\ \ref{fig_fund2}
with results of Monte Carlo simulations. One can see a rapid convergence and
already for $n=4$ the difference between the simulation and the cluster
result is extremely small.

\subsection{Car-Oriented Mean-Field Theory}
\label{seccomf}

The car-oriented mean-field (COMF) theory \cite{COMF} is another possibility 
to take into account correlations in an analytical description\footnote{A
similar method for reaction-diffusion models is discussed in 
\cite{interpart}.}.

The rules R1--R4 can be rewritten in terms of $d_j$ and $v_j$ only. Therefore
the state of the system can be characterized instead of the occupation
numbers $\tau_j$ equivalently by the headways and the 
velocities\footnote{For a complete equivalence one has to track the position
$x_1$ of one car, e.g.\ car 1. From $x_1$, $\{d_j\}$ and $\{v_j\}$ one
can then determine $\{\tau_j\}$.}.
The central quantitiy in COMF is the probability $P_n(v)$ to find 
exactly $n$ empty cells (i.e.\ a gap of size $n$) in front of a car 
with velocity $v$. In this way certain longer-ranged correlations
are already taken into account. The essence of COMF is now to neglect 
correlations between the headways.

Using again the update ordering R2-R3-R4-R1 (see Sec.\ \ref{sec_clust}),
one obtains for $\vm=1$ the following system of equations resulting from 
the master equation for the stationary state (with $P_n=P_n(v=1)$):
\begin{eqnarray}
P_0&=& \bar g\left[P_0+\q P_1\right]\ ,\nonumber\\
P_1&=& gP_0 + \left[\q g+p\bar g\right]P_1 
          + \q \bar gP_2\ ,\label{comfv1}\\
P_n&=& pgP_{n-1} + \left[\q g+p\bar g\right]P_n
+ \q \bar gP_{n+1}\ ,\quad  (n\geq 2)\nonumber
\end{eqnarray}
where $g=\q \sum_{n\geq 1} P_n = \q [1-P_0]$ 
is the probability that a car moves in the next timestep. 
$\bar g=1-g$ then is the probability that a car does not move.

As an example for the derivation of these equations we consider the
equations for $n\geq 2$. Since the velocity difference of two cars is at
most 1, a gap of $n$ cells at time $t+1$ must have evolved form a gap
of length $n-1$, $n$, or $n+1$ in the previous timestep. A headway of $n-1$
cells evolves into a headway of $n$ cells only if the first car moves (with
probability $g$) and the second car brakes in the randomization step
(probability $p$), i.e.\ the total probability for this process is 
$pgP_{n-1}$. Similarly, the headway remains constant only if either
both cars move (probability $\q g$) or both cars do not move (probability 
$p{\bar g}$). Finally, the headway is reduced by one, if only the
second car moves (probability $\q {\bar g}$).

The probabilities $P_n$ have to satisy the normalization conditions
\begin{eqnarray}
1 &=&  \sum_{n=0}^\infty P_n\ , \label{COMFnorm}\\
\frac{1}{c} &=& \sum_{n=0}^\infty (n+1)P_n\ .\label{COMFdensity}
\end{eqnarray}
(\ref{COMFdensity}) is a consequence of the conservation of the number
of cars since each car with headway $n$ occupies $n+1$ cells.

Although the infinite system of non-linear equations (\ref{comfv1})
looks more difficult than those of the cluster approximation, a solution 
possible using generating functions \cite{COMF}. For $\vm=1$ one finds
\begin{eqnarray}
P_0 &=& \frac{1}{2\q c}\left[2\q c-1+\sqrt{1-4\q c(1-c)}\right]\ ,\nonumber\\
P_n &=&\frac{P_0}{p}\left(\frac{p(1-P_0)}{P_0+p(1-P_0)}\right)^n
\quad (n\geq 1),
\label{gapdist}
\end{eqnarray}
and for the flow $f(c,p)=cg$ again the exact solution is reproduced.

For $\vm=2$ one has two coupled systems of the type (\ref{comfv1}), since 
one has to distinguish $P_n(v=1)$ and $P_n(v=2)$ and the probabilities
$g_\alpha$ that a car moves $\alpha=0,1,2$ cells in the next timestep. 
This system has a similar structure as (\ref{comfv1}) and is given 
explicitely in \cite{COMF}. It can also be solved, but does not 
give the exact solution.

Note that the COMF approach assumes an infinite system size. For a finite
system of length $L$ the largest headway that can appear is $M:=L-N$.
For $\vm=1$ it is possible to satisfy (\ref{COMFnorm}) and (\ref{COMFdensity})
by just changing the equations for $P_{M-1}$ and $P_{M}$:
\begin{eqnarray}
P_{M-1}&=& pgP_{M-2} + \left[\q g+p\bar g\right]P_{M-1} + 
\left[\q \bar{g}-p g\right]P_{M}\ ,\nonumber\\
P_{M}&=& pgP_{M-1} + \left[1-\q \bar{g}+p g\right]P_{M}\ .
\label{COMFfinite}
\end{eqnarray}
The equations for $P_n$ with $n< M-1$ are the same as in (\ref{comfv1}).
In order to derive (\ref{COMFfinite}) we have made the Ansatz
$P_{M-1}=\alpha_1P_{M-2}+\alpha_2P_{M-1}+\alpha_3P_{M}$ and
$P_{M}=\beta_1P_{M-1}+\beta_2P_{M}$ and determined the cofficients
$\alpha_j$ and $\beta_j$ from (\ref{COMFnorm}) and (\ref{COMFdensity}).

For $\vm=2$ it appears that the equations for all $n$ have to be modified
in order to satisfy (\ref{COMFnorm}) and (\ref{COMFdensity}). This
corresponds nicely to the qualitative difference of the cases $\vm=1$ and
$\vm>2$ and the appearance of long-ranged correlations in the latter
(see Sec.\ \ref{sec_GoE}).


\subsection{Garden of Eden States}
\label{sec_GoE}

\begin{figure}[ht]
\centerline{\psfig{figure=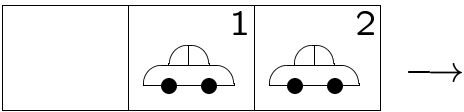,bbllx=150pt,bblly=435pt,bburx=460pt,bbury=520pt,height=2cm}}
\caption{A Garden of Eden state for the model with $\vm \geq 2$.}
\label{fig_GoEconf}
\end{figure}

An important effect of the parallel dynamics is the existence of 
configurations which can not be reached dynamically \cite{eden}. These 
states are called Garden of Eden (GoE) states or paradisical states since 
one never gets back once one has left \cite{goe}. 
An example for a GoE state is given in
Fig.\ \ref{fig_GoEconf}. Note that the velocity is equal to the number 
of cells that the car moved in the previous timestep. For the 
configuration shown in Fig.\ \ref{fig_GoEconf} this implies that the two 
cars must have occupied the same cell before the last timestep.
Since this is forbidden in the NaSch model, the configuration shown can 
never be generated by the dynamics.

The simple mean-field theory presented in the previous section does not
take into account the existence of GoE states. One can therefore hope
that by eliminating all GoE states and applying mean-field theory in
the reduced configuration space (paradisical mean-field, pMF) 
one will find a considerable improvement of the MF results.

For $\vm=1$ all states containing the local configurations
$(0,1)$ or $(1,1)$ are GoE states, i.e.\ the cell behind a car with velocity 1
must be empty\footnote{Here we do not use the changed update order, so '0'
denotes a cell occupied by a car with velocity 0.}. 
This only affects eq.\ (\ref{mf1}) and the equations for pMF theory read:
\begin{eqnarray}
c_0&=&\nn (c_0+pd)c\ ,\label{pmf10}\\
c_1&=&\nn \q cd\ ,\label{pmf11}
\end{eqnarray}
where the normalization $\nn$ ensures $c_0+c_1=c$ and is given
explicitly by $\nn =1/(c_0+d)$. 

Solving (\ref{pmf11}) for $c_1$ by using $c_0=c-c_1$, one obtains
$c_1$ as a function of the density $c$:
\begin{equation}
c_1 = \frac{1}{2}\left(1-\sqrt{1-4\q (1-c)c}\right)
\label{GoEc1}
\end{equation}
(with $\q =1-p$).  Since the flow is given by $f(c)=c_1$
we recover the exact solution for the case $\vm=1$ (see Section
\ref{sec_clust}). 

\begin{figure}[t]
\centerline{\psfig{figure=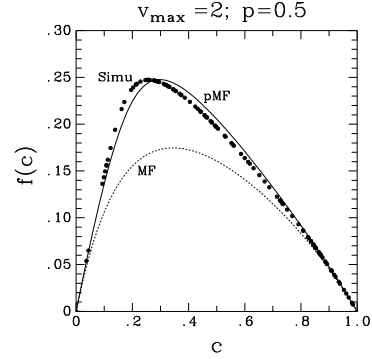,bbllx=55pt,bblly=160pt,bburx=570pt,bbury=680pt,height=5cm}}
\caption{Comparison of the fundamental diagrams obtain from MC simulations
with the results of mean-field theory (MF) and mean-field theory without
GoE states (paradisical mean-field, pMF).}
\label{fig_GoE}
\end{figure}

It is interesting to note that for random-sequential dynamics MF theory
is exact whereas for parallel dynamics pMF is exact. 
Therefore the origin of the correlations is the parallel update
procedure. The existence of the GoE states is responsible for the 
differences between parallel and random-sequential dynamics.
This is probably not only true for the NaSch model, but is a rather 
general property of CA models.

For $v_{max}=2$ one has to take into account more GoE states \cite{eden}.
pMF is no longer exact, but it still leads to a considerable 
improvement of the MF results (see Fig.\ \ref{fig_GoE}).


\section{``Physical'' Quantities}
\label{sec_phys}

In this section we will use the analytical methods of 
Sec.\ \ref{sec_approx} to determine physical quantities of the
NaSch model, e.g.\ the fundamental diagram, cluster-size distributions
and correlation functions. These results will be used in 
Sec.\ \ref{sec_struct} to gain a better understanding of the microscopic
structure of the stationary state.


\subsection{Fundamental Diagram}
\label{sec_funda}

For $\vm=1$ the fundamental diagram in MF approximation is given by
\begin{equation}
f_{MF}(c)=\q c(1-c)\ .
\end{equation}
As already mentioned in this case 2-cluster approximation, COMF and
pMF are exact and the flow is given
\begin{equation}
f(c,p) = \frac{1}{2}\left(1-\sqrt{1-4\q (1-c)c}\right)
\label{v1flow}
\end{equation}
(with $\q =1-p$). The fundamental diagram is symmetric with respect to
$c=1/2$, reflecting the particle-hole symmetry of the NaSch model with
$\vm=1$. 

It is apparent from Fig.\ \ref{fig_vmax1} that MF considerably underestimates
the flow for intermediate densities. 
This shows that correlations are important in this regime. These
correlations lead to an increase of the flow. One finds a
particle-hole attraction (particle-particle repulsion), i.e.\ the 
probability to find an empty site in front of a car is enhanced 
compared to a completely random configuration. 
In terms of cluster probabilities this particle-hole attraction can be
expressed more quantitatively as $P(1,0) > P(1)P(0)=c(1-c)$.

As mentioned in Sec.\ \ref{sec_GoE}, pMF yields the exact solution for
$\vm=1$. This shows that there are no 'true' correlations apart from those 
due to the existence of GoE states. The 2-cluster approximation and COMF
then become exact since both methods are able to identify all GoE states.

For $\vm >1$ one expects stronger correlations than for $\vm=1$ \cite{SSNI}.
Indeed for $\vm=2$ the difference between the MF prediction 
\begin{equation}
f_{MF}(c)=c_1+2c_2=\frac{\q (1+\q d^2)dc}{1-pd^2}
\end{equation}
and the Monte Carlo data is even larger (see Fig.\ \ref{fig_fund2}). However,
pMF is not exact and therefore 'true' correlations exist. This is in 
agreement with the fact that MF is also not exact for the NaSch model
with $\vm>1$ and random-sequential dynamics.

The correlations can be taken into account systematically by the $n-$cluster 
approach which converges rapidly with increasing $n$. Already for $n=4$ one
obtains a very good agreement with the simulation results
(Fig.\ \ref{fig_fund2}). The quality of COMF depends strongly on the value
of $p$. In general its prediction is between the 2-- and 3--cluster 
approximation.


\subsection{Headway Distribution}

\begin{figure}[ht]
\centerline{\psfig{figure=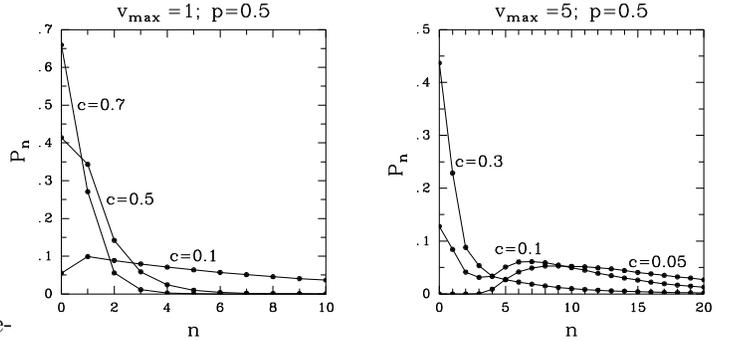,bbllx=40pt,bblly=420pt,bburx=580pt,bbury=685pt,height=4.8cm}}
\caption{Distribution of headways for different densities for $\vm =1$ 
and $p=0.5$  (left) and $\vm =5$ and $p=0.5$ (right).}
\label{fig_COMF}
\end{figure}

Fig.\ \ref{fig_COMF} shows the result for the distribution of headways, 
$P_n$, for $\vm=1$. Since in this case COMF is exact, we can use the
result (\ref{gapdist}). The headway distribution has just one maximum,
located at $n=1$ for small densities and at $n=0$ for large $c$. 

These two regimes also exist for higher velocities. At high densities the
headway distribution is maximal at $n=0$, whereas for low densities the
maximum is found at some value $n_{max}>0$, where $n_{max}$ depends on the
density. In addition, an intermediate density regime exists, where 
distribution exhibits two local maxima, one at $n=0$, corresponding to
jammed cars, and one at $n_{loc}>0$, corresponding to free flowing 
cars \cite{chowd1}.

\subsection{Correlation Length}

As an application of the cluster approximation we will compute 
density-density correlations for $\vm =1$ in the following. Using occupation
numbers $n_j=0,1$ the density-density correlation function is 
defined by
\begin{equation}
\langle n_1 n_r\rangle = {\sum_{\{n_j\}}}'n_1n_r P(n_1,\ldots,n_L)
\label{corrdef}
\end{equation}
where the prime indicates that the sum runs over all states with fixed 
particle number $N=n_1+\cdots +n_L$.

In order to evaluate the sum in (\ref{corrdef}), it is convenient to use
a grand-canonical description. One introduces a fugazity $z$ which controls
the average number of particles $\langle n_j\rangle$ and sums over all
configurations in (\ref{corrdef}). Using the 2-cluster approximation
(\ref{2clusterapp}) -- which is exact for $\vm=1$ -- the correlation 
function is given by
\begin{eqnarray}
\langle n_1 n_r\rangle &=& \frac{1}{Z_{gc}}{\sum_{\{n_j\}}} n_1n_r 
z^{N} P(n_1,n_2)P(n_2,n_3)\cdots\nonumber\\
&&\qquad\qquad\cdots P(n_{L-1},n_L)P(n_L,n_1)
\label{corrgc}
\end{eqnarray}
with $N=\sum_{j=1}^L n_j$ and the normalization
\begin{equation}
Z_{gc}=\sum_{\{n_j\}}z^{N}\prod_{j=1}^L P(n_j,n_{j+1}).
\end{equation}
Introducing the transfer matrix
\begin{equation}
{\Pt}=\begin{pmatrix} 
P(0,0)         & {\sqrt{z}}P(0,1) \\
{\sqrt{z}}P(1,0) & zP(1,1)
\end{pmatrix}
\end{equation}
this can be written succinctly as
\begin{eqnarray}
Z_{gc}&=&{\rm Tr\ } {\Pt}^L\ ,\\
\langle n_1 n_r\rangle &=&\frac{1}{Z_{gc}}\, {\rm Tr}\left({\mathbf Q}
\Pt^{r-1}{\mathbf Q}{\Pt}^{L-r+1}\right)\ .
\label{corrlength}
\end{eqnarray}
with the matrix $Q(n_1,n_2)= n_1\Pt(n_1,n_2)$.

The correlation length $\xi$ can be obtained from the asymptotic 
behaviour ($r\to\infty$) of the correlation function 
\begin{equation}
\langle n_1 n_r\rangle -c^2 \propto e^{-r/\xi}
\end{equation}
where $c=\langle n_j\rangle$ is the (average) density of cars. 
$\xi$ is determined by the ratio 
of the eigenvalues $\lambda_\pm$ of $\Pt$ (with $|\lambda_+| \geq 
|\lambda_-|$):
\begin{equation}
\xi^{-1}=\ln\left|\frac{\lambda_-}{\lambda_+}\right|\ .
\end{equation}
The explicit expression for $\xi$ can be obtained from
\begin{eqnarray}
\lambda_\pm &=& \frac{1}{2}\left[ A\pm \sqrt{A^2+4\q z[P(1,0)]^2}\right]
\ ,\\
{\rm with\ \ \ } A &=& P(0,0)+zP(1,0)\ .
\end{eqnarray}
$P(a,b)$ are the cluster probabilities given in (\ref{2clustsol1}).
The fugazity $z$ can be related to the density $c$ via the equation
$c=\langle n_j \rangle$. Using an expression analogous to (\ref{corrlength})
one finds
\begin{eqnarray}
c\lambda_+&=&\frac{B\left\{[P(1,0)]^2+BP(1,1)\right\}}{[P(1,0)]^2
+\frac{1}{z}B^2}\ , \\
{\rm with\ \ \ } B&=& \lambda_+-P(0,0)\ .
\end{eqnarray}

For fixed $p$, $\xi$ is maximal for $c=1/2$ which corresponds to
$z=1$.  $\xi(c=1/2)$ diverges only for $p\to 0$. In that case one
finds that $\xi(c=1/2)\propto p^{-1/2}$.  Monte Carlo simulations show
that the same behaviour still occurs (at density $c=1/(\vm+1)$)
for $\vm >1$ \cite{eisi}.  Therefore, the correlation function already
gives an indication that the system is not critical for $p>0$. The
simulations show that there is no qualitative difference between the
cases $\vm=1$ and $\vm >1$ as far as the phase transition is concerned
\cite{eisi}.

The above results demonstrate that, although the 2-cluster approximation
is exact, not all correlation functions are of finite range.


\subsection{Jam-Size Distribution}
\label{sec_jamsize}

For a better understanding of the differences between the cluster
approximation and COMF we calculate the distribution of jam-sizes using these
methods. Let $C_n $ be the probability to find a (compact) jam of 
length $n$, i.e.\ $n$ consecutive occupied cells.

In COMF, $C_n$ is proportional to $(1-P_0)$$P_0\cdots P_0(1-P_0)=(1-P_0)^2
P_0^{n-1}$. The number of jams is proportional to $1-P_0$ so that one finds
\begin{equation}
  C_{n}=(1-P_0)P_0^{n-1}\ .
\label{COMFjamdist}
\end{equation}
This distribution is purely exponential and $P_n \geq P_{n+1}$ for all $n$.
Therefore COMF is not able to describe clustering or phase separation,
i.e.\ situations where jams with more than one car dominate. Clustering
implies that there are correlations between gaps which are completely
neglected in COMF.

In 2-cluster approximation, $C_n$ is proportional to 
$P(\underline{0}|1)$ $P(\underline{1}|1)\cdots$ 
$P(\underline{1}|1)P(\underline{1}|1) P(\underline{1}|0)$ where we have
used the conditional probabilities (\ref{condprob2}).
and the different update-ordering R2-R3-R4-R1.  Note that
the case $n=1$ has to be treated separately, $C_1 \propto\sum_{v=1}^{\vm}
P(\underline{0}|v) P(\underline{v}|0)$. The number of jams is obviously 
given by ${\cal N}_J=\sum_{v=1}^{\vm} P(\underline{v}|0)$ and one obtains
\begin{eqnarray}
C^{(2)}_n &=& \frac{1}{{\cal N}_J}\, P(\underline{0}|1) 
    P(\underline{1}|1)^{n-2}P(\underline{1}|1) P(\underline{1}|0)
\quad (n\geq 2), \nonumber\\
C^{(2)}_1 &=& \frac{1}{{\cal N}_J}\sum_{v=1}^{\vm} P(\underline{0}|v) 
    P(\underline{v}|0)\ .
\end{eqnarray}
$C^{(2)}_n$ decays exponentially for $n\geq 2$, especially one has
$C_{n+1}\leq C_n$.
For the $m$-cluster approximation one can derive similar expressions. 
Since here clusters of size $m$ are treated exactly it is now possible
to find the maximum of the jam-size distribution somewhere between
$n=1$ and $n=m$. For $n>m$ the jam-size probability decays exponentially
due to the mean-field-like self-consistent coupling of the cluster to
the rest of the system.



\subsection{Further Applications}

We briefly mention other results which have been obtained analytically
using the cluster approximation. In \cite{chowd1} the distributions of
gaps and the distance between jams have been calculated for $\vm=1$ using
the 2-cluster approximation. For the gap distribution one recovers the
(exact) result which has been derived in Sect.\ \ref{seccomf} (see 
eq.\ (\ref{gapdist})). For the calculation of the distribution of gaps between
jams one defines every vehicle with velocity 0 to be jammed. The distance
between two jams is then given by the distance between a car with
velocity 0 and the next car with velocity 0.
In \cite{chowd2} the probability $\cP(t)$ of a time headway $t$ has been
investigated. This quantity is defined in analogy to measurements on
real traffic where a detector registers the time interval between
the passing of consecutive cars.
For a discussion of the behaviour of the time-headway and the jam-gap
distribution in the NaSch model we refer to \cite{chowd1,chowd2}. These
quantities are also very useful for studying variants of the NaSch model
which exhibit phase separation \cite{debchs2s}.


\section{Microscopic Structure of the Stationary State}
\label{sec_struct}

In the following  we demonstrate that the microscopic structure of the 
stationary state changes qualitatively with the braking parameter $p$.
For illustration we first discuss the deterministic limits $p=0$ and
$p=1$. After that the behaviour for randomizations $0<p<1$ is discussed.
However, we concentrate on the limits $p\ll 1$ and $1-p \ll 1$.
Finally, in Sec.\ \ref{sec_vs} we compare COMF and cluster approximation.
Here we focus on the ability to reproduce the correct microscopic
structure of the stationary state.


\subsection{$\mathbf{p=0}$}
\label{sec_p0}

In the absence of random decelerations the velocity of the vehicles
is determined solely by its headway. For densities $c \leq 1/(\vm +1)$
all headways can be larger than $\vm$ and therefore the cars move with
velocity $\vm$ in the stationary state. 
This is no longer possible for $c > 1/(\vm +1)$. Here the velocity
of the cars is determined by the average headway $\bar{d}$ available, 
$\bar{d}=(L-N)/N=(1-c)/c$.
The fundamental diagram is therefore given by
\begin{equation}
f(c)=
\begin{cases}
\vm c & \text{for $c \leq 1/(\vm +1)$,}\\
1-c & \text{for $c> 1/(\vm +1)$.}
\end{cases}
\end{equation}
Note that in general the stationary state is not unique, but determined 
completely by the initial condition.

For $p=0$ there is no tendency towards clustering \cite{nagelherrm}. 
Overreactions are not possible and there is no spontaneous formation 
of jams. In this sense the case $p=0$ is unrealistic. The dynamics is 
completely determined by ``geometrical'' effects, since the behaviour 
of the cars only depends on the available average headway. There is no 
'attractive' interaction between the cars and therefore no mechanism 
for clustering.



\subsection{$\mathbf{p=1}$}
\label{sec_p1}

In the case $p=1$ the dynamics is again deterministic. However, it is
very different from that of a model with maximum velocity $\vm-1$. The
point that we want to stress here is the existence of metastable states.
For $\vm=2$ and density $c=1/3$ the state $..1..1..1..$ (here '.' denotes
an empty cell and '1' a cell occupied by a car with velocity 1) is
stationary with flow $f(c=1/3)=1/3$. On the other hand, the state
$..0..0..0..$ is also stationary with vanishing flow, since a standing
car will never start to move for $p=1$.

For densities $c>1/3$ all stationary states have vanishing flow. Here
at least one car has only one empty cell in front. Therefore after step R2
this car has velocity 1 and will then decelerate to velocity 0 in step R3.
After that it will never move again.

For densities $c\leq 1/3$ stationary states with non-zero flow exist.
These are not stable under local perturbations, i.e.\ stopping just one 
car leads to a complete breakdown of the flow. In this sense these states
are metastable.

Starting from random initial conditions, the metastable states have a
vanishing weight in the thermodynamic limit, since already one standing
car leads to a zero-flow state.

Note that the metastable states at $p=1$ do not exist for $\vm =1$. In
that sense the difference between $\vm=1$ and $\vm>1$ becomes most
pronounced at $p=1$.

The deceleration introduces a kind of 'attractive' interaction between
the cars which can lead to the formation of jams. However, these jams
are typically not compact, but of the form $.0.0.0.0.$.
A car approaching a standing car adapts its velocity in step R2 such that
it reaches the cell just behind the standing vehicle. In step R3 it then
decelerates further and so there will be a gap of 1 between the cars.


\subsection{$\mathbf{0<p<1}$}
\label{sec_p0to1}

\begin{figure}[ht]
\centerline{\psfig{figure=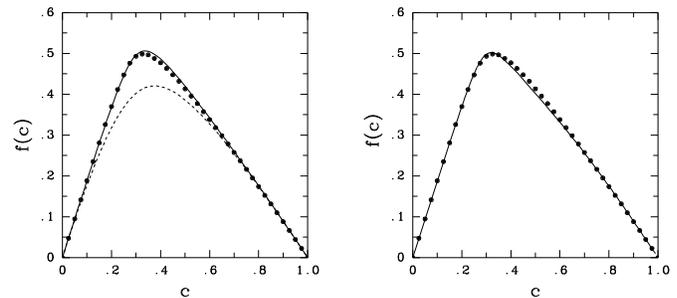,bbllx=45pt,bblly=415pt,bburx=575pt,bbury=695pt,height=4.7cm}}
\caption{Fundamental diagram for $\vm=2$ and $p=0.1$: Comparison of 
simulation data ($\bullet$) with the 2-cluster ($\cdots$) and 3-cluster 
results ({\bf ---}) (left) and with COMF ({\bf ---}) (right).}
\label{fig_smallp}
\end{figure}

In the limit $p\to 0$, the fundamental diagrams obtained from COMF and 
the 3-cluster approximation become exact, in contrast to the 2-cluster 
approximation (see Fig.\ \ref{fig_smallp}). Even for values of $p\approx 0.1$
there is an excellent agreement between the fundamental diagrams obtained
analytically and the Monte Carlo simulations. Since 'realistic' values
of $p$ are in the region $p \sim 0.1 - 0.2$, the approximations are
indeed applicable in the relevant parameter regime.

\begin{figure}[ht]
\centerline{\psfig{figure=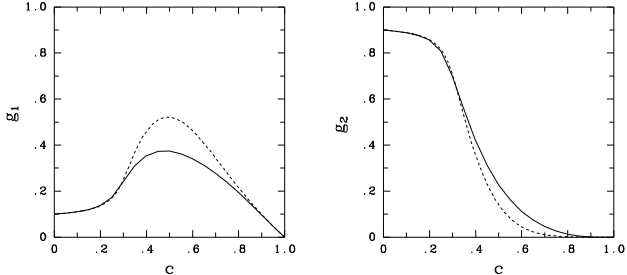,bbllx=40pt,bblly=410pt,bburx=570pt,bbury=670pt,height=4.1cm}}
\caption{Comparison of the COMF prediction ($\cdots$) with 
computer simulations ({\bf ---}) for 
the probabilities $g_\alpha$ that a car moves $\alpha$ sites.
$g_0$ can be obtained from $g_0=1-g_1-g_2$. The braking parameter is
$p=0.1$ and $\vm=2$.}
\label{fig_g1g2}
\end{figure}

Fig.\ \ref{fig_g1g2} shows a comparison of the computer simulations with the
COMF results for the probabilities $g_\alpha$ that a car moves $\alpha$ sites.
Although there is an excellent agreement for the fundamental diagram
$f(c)=c(g_1+2g_2)$ at $p=0.1$ (see Fig.\ \ref{fig_smallp}), there are
considerable deviations at intermediate densities for the individual
values of $g_1$ and $g_2$. This is somewhat surprising and indicates that
COMF is not exact in the limit $p\to 0$, i.e.\ that there are still 
correlations between the headways. This will be discussed further in
Sec.\ \ref{sec_vs}.

\begin{figure}[t]
\centerline{\psfig{figure=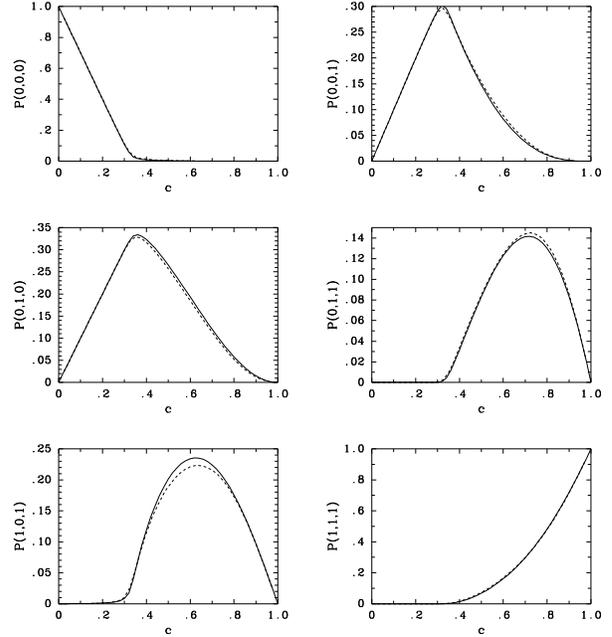,bbllx=35pt,bblly=190pt,bburx=580pt,bbury=790pt,height=9cm}}
\caption{Comparison of the 3-cluster probabilities $P(n_j,n_{j+1},n_{j+2})$
obtained from computer simulations ({\bf ---}) and the 3-cluster 
approximation ($\cdots$) for $p=0.01$.
$n_j$ is the occupation number of cell $j$.
Note that $P(1,0,0)=P(0,0,1)$ and $ P(1,1,0)=P(0,1,1)$.}
\label{fig_P3}
\end{figure}

On the other hand, the 3-cluster probabilities obtained from our simulations
show a very good agreement with the results of the 3-cluster approximation
(Fig.\ \ref{fig_P3}). This suggests that the 3-cluster approximation is
asymptotically exact for $p\to 0$.

\begin{figure}[t]
\centerline{\psfig{figure=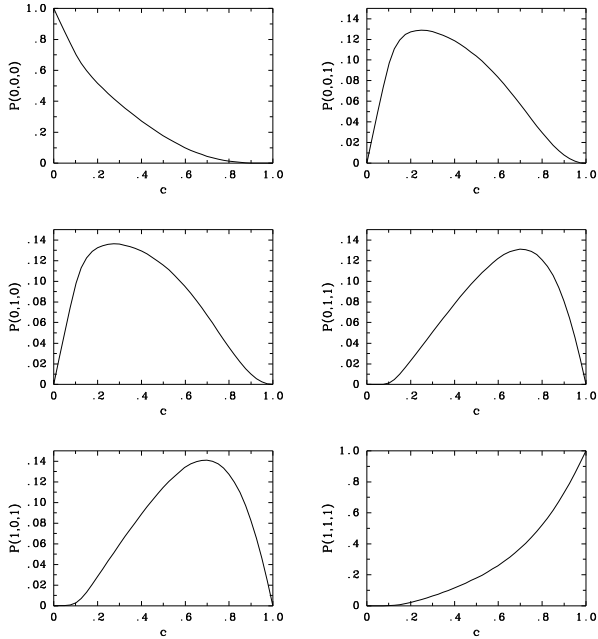,bbllx=35pt,bblly=190pt,bburx=580pt,bbury=790pt,height=9cm}}
\caption{Same as Fig.\ \ref{fig_P3}, but for $p=0.75$. Here only the results
of the computer simulations are shown.}
\label{fig_P3_75}
\end{figure}

The limit $p\to 1$ is difficult to investigate numerically. Preliminary
results show that COMF is not exact in this limit. The cluster
approximations give a much better agreement with Monte Carlo simulations,
but it is not clear yet whether it becomes exact or not. 
However one can expect a tendency towards phase seperation due to
'attractive' interactions beween the cars. One stopped car can induce
jams with a rather long lifetime since the restart probability of the
first car is rather small. This tendency can already be observed
in the cluster probabilities for $p=0.75$ (see Fig.\ \ref{fig_P3_75}).
The distributions become broader and especially at small densities
the probabilities $P(0,1,1)$ and $P(1,1,1)$ which characterize the
clustering are enhanced compared to the limit $p\to 0$.


\subsection{Cluster Approximation vs.\ COMF}
\label{sec_vs}

In this section the results of the cluster approximation and COMF in special 
limits are compared. As already mentioned, both methods yield the exact
solution for $\vm=1$. This is related to the fact that paradisical mean-field
is exact in that case and both methods are able to identify all
GoE states. For $\vm=2$ the situation is different. Here all three
methods are only approximative. 

The deviations between COMF results and the Monte Carlo simulations
can be understood by looking more closely at the microscopic structure
of the stationary state. For $p=0$ there are three different stable
structures: $00000$, $.1.1.1.$, and
$..2..2....2...2...$. The last structure comprises all configurations
where all cars have velocity $\vm$ and at least $\vm$ empty cells in front.
For fixed density there are many configurations
which produce the same flux. In a large system these configurations are
mostly made up of combinations of the three stable structure (with small
transition regions). For $p>0$, $.1.1.1.$ is the most unstable
configuration. It tends to seperate into $00000$ and 
$2..2....2...2$ under fluctuations of the headways.
COMF is not able to account for these fluctuations since headways are
assumed to be independent. Therefore the weight of the configurations
$.1.1.1.$ -- and therefore $g_1$ -- is overestimated.
In the 3-cluster approach, on the other hand, the headways are not
independent. Therefore it is able to identify the dominating local
structure for $p\to 0$ correctly.

COMF is not able to describe the 'attractive' part of the interaction
properly. This can also be seen in the jam-size distribution (see
Sec.\ \ref{sec_jamsize}). COMF always predicts a strictly monotonous
distribution (\ref{COMFjamdist}), whereas the cluster approximation
in principle is able to describe situations where the jam-size distribution
has a maximum at some (small) value of $n>0$.


\section{Summary}
\label{sec_summary}

Although cellular automata are designed for  efficient computer simulation
studies, an analytical description is possible, although difficult.
We have presented here four different methods which can be applied
to CA models of traffic flow. The first approach, a simple mean-field
theory for cell occupation numbers, is insufficient since the important 
correlations between neighbouring cells (e.g.\ the particle-hole attraction) 
are neglected. We therefore suggested three different improved mean-field
theories. These approaches take into account certain correlations 
between the cells. The simplest method is the so-called 'paradisical
mean-field' theory which is based on the observation that certain
configurations (Garden of Eden states) can never be generated by the
dynamical rules due to the use of parallel dynamics. The cluster
approximation, on the other hand, treats clusters of a certain size
exactly and couples them in a self-consistent way. Therefore short-ranged
correlations are taken into account properly. In contrast, car-oriented 
mean-field theory is a true mean-field theory, but here one uses a 
different dynamical variable, namely the distance between consecutive
cars. In that way, certain correlations between cells are taken into
account.

All three improved MF theories become exact for $\vm=1$. For larger
values of $\vm$ they are just approximations. In principle, the cluster
approximation and COMF (in combination with a cluster approach) can be
improved systematically. This is, however, very cumbersome.

An interesting observation is that the qualitity of the approximation 
depends strongly on the value of $p$. This indicates that the physics 
changes with $p$, contrary to common believe. Evidence for this
scenario comes from the microscopic structure of the stationary state.
In the limit $p\to 0$ it is dominated by repulsive interactions which
tend to aligne the vehicles at a headway of at least $\vm$ cells.
On the other hand, for $p\to 1$ there is a tendency towards
phase seperation. Cars which had to stop due to a fluctuation will stand 
for a rather long time and thus lead to the creation of jams. In the
deterministic case this even leads to the existence of metastable states.

In \cite{janos} it has been suggested that the behaviour of the
NaSch model is governed by two fixed points,
namely $p=0$ and $\vm=\infty$. Our investigations show that
it might be more reasonable to consider $p=0$ and $p=1$ as fixed points 
in order to understand the behaviour for fixed $\vm$.
For $p=0$ there is a continous phase transition from laminar flow
to a congested phase (see \cite{eisi,janos} and references therein)
at $c=1/(\vm+1)$. This
transition turns into a crossover at $0<p<1$ \cite{eisi,janos} . It
would be desirable to investigate the behaviour of this transition close
to $p=1$ in more detail.

The methods presented here can also be used for other CA models,
e.g. variants of the NaSch model 
\cite{chowd2,debchs2s,slow2,wang1,wang2,ASrev}.
The investigation of the NaSch model has led to a better understanding
of their advantages and limitations so that it is easier to choose
the approach most suitable for a given problem.

%
\paragraph{Acknowledgments}

I would like to thank Michael Schreckenberg, Ludger Santen, Debashish
Chowdhury and Dietrich Stauffer for valuable discussions and useful
suggestions. This work has been performed within
the research program of the SFB 341 (K\"oln--Aachen--J\"ulich).

\newpage
%
%

\end{document}